\documentclass[12pt,draftclsnofoot, onecolumn,draftcls]{IEEEtran}
\usepackage{graphicx,cite,epsfig,amssymb,amsmath,subfigure,url,stfloats,latexsym}
\usepackage{multirow,array}
\usepackage{graphicx,cite,amsmath,amssymb,hhline,booktabs,stfloats,bm}
\usepackage{verbatim}
\usepackage[utf8]{inputenc}
\usepackage{subfigure}
\usepackage{algorithm,algorithmic}
\usepackage{color}

\begin{document}
\setcounter{page}{1}

\title{\mbox{}\vspace{1.5cm}\\
\textsc{
Matching Model of Flow Table for Networked Big Data} \vspace{1.5cm}
}

\author{Yiheng Su, Ting Peng, Xiaoxun Zhong, and Lianming Zhang
\thanks{Yiheng Su, Ting Peng, Xiaoxun Zhong, and Lianming Zhang  (zlm@hunnu.edu.cn) are with Key Laboratory of Internet of Things Technology and Application, College of Physics and Information Science, Hunan Normal University, Changsha, China (Corresponding Author: L. Zhang).}
}

\date{\today}
\renewcommand{\baselinestretch}{1.2}
\thispagestyle{empty} \maketitle \thispagestyle{empty}

\begin{abstract}
Networking for big data has to be intelligent because it will adjust data transmission requirements adaptively during data splitting and merging. Software-defined networking (SDN) provides a workable and practical paradigm for designing more efficient and flexible networks. Matching strategy in the flow table of SDN switches is most crucial. In this paper, we use a classification approach to analyze the structure of packets based on the tuple-space lookup mechanism, and propose a matching model of the flow table in SDN switches by classifying packets based on a set of fields, which is called an F-OpenFlow. The experiment results show that the proposed F-OpenFlow effectively improves the utilization rate and matching efficiency of the flow table in SDN switches for networked big data.

\end{abstract}

\begin{IEEEkeywords}
\begin{center}
Networked Big Data, Software-Defined Networking, Matching Model, F-OpenFlow, Tuple-Space Search
\end{center}
\end{IEEEkeywords}

\IEEEpeerreviewmaketitle

\vspace{0.3in}
\section{Introduction}

With the advent of big data \cite{b1}, networking for big data, also called networked big data, will become more and more important \cite{b2}. The applications with bigger and bigger data volume need to process vast information, whose scale grows larger and more complex to the multiple locations for resiliency. Networked big data needs an intelligent network, which can adaptively adjust data transmission requirements during data splitting and merging, thus achieving a fast run speed and high utilization ratio. Software-defined networking (SDN) \cite{b3} is a new networking innovation paradigm that separates the control plane from the data plane, offering a well-defined programmable interface for software intelligence. The SDN can meet the requirements of networked big data through the network programming, making it possible to analyze networked big data, and build a self-adaptive intelligent network. In the recent years, Hu \emph{et al}. proposed an SDN-enabled big data processing system for social TV analytics \cite{b4}. Cui \emph{et a}l. proposed dynamic traffic engineering system architecture and cross-layer design with SDN and big data.\cite{b5}. Lin \emph{et al}. presented a forward-looking view of the convergence of IoT, big data, cloud and SDN technologies along with the arrival of 5G mobile broadband networks \cite{b6}. Qin \emph{et al}. \cite{b7} proposed a heuristic bandwidth-aware for implementing task scheduler bandwidth-aware scheduling with SDN in Hadoop for big data.

The two main components of SDNs are SDN controllers and SDN switches. The SDN controllers are mainly responsible for the centralized management and remote control of network devices, while the SDN switches relying on flow tables are only in charge of forwarding data. However, the global deployment of SDNs encounters some issues. With the increasing number of flow tables and the complicated structure of flow entries, the size of flow tables in SDN switches explosively grows, which poses difficulties for practicing flow tables by hardware \cite{b8}. To solve the above questions, the authors of paper \cite{b8} provided a model for minimizing storage space of flow tables and proposed an H-SOFT algorithm. Guo \emph{et al}. \cite{b9} proposed a forwarding scheme for achieving low and balanced usage of flow tables by properly and reactively placing flow entries in SDN switches. Luo \emph{et al}. proposed fast flow table aggregation (FFTA) and its incremental FFTA to make practical flow table aggregation \cite{b10}. Huang \emph{et al}. \cite{b11} proposed a rule partition and allocation algorithm to distribute the rules across network switches. Leng \emph{et al}. \cite{b12} proposed a reduction scheme to solve flow table congestion problem. Metter \emph{et al}. \cite{b13} formulated a simple analytical model that allows optimizing the network performance with respect to the flow table occupancy.

The Open vSwitch kernel has been adopted Microflow Cache method \cite{b14}, where the Hash's exact lookup table and wildcard lookup table they use it to reduce the number of lookups in a multilayer flow table. Since Open vSwitch utilizes a tuple-space lookup method, the average number for searching the original lookup table is half of what adopts the tuple-space table based on the kernel. The core of this idea is to minimize the number of messages in the lookup of multilayer flow tables, and to increase the number of accesses to the lookup table in the cache of switches.

In this paper, we present a matching model of the flow table in SDN switches by classifying packets based on matching a set of fields, which is short for F-OpenFlow, to search flow tables of OpenFlow switches. The main contributions of this paper are as follows:

\begin{itemize}
\item Proposing the method that the matching field can improve the matching probability of the table entries by classifying packets;
\item Using tuple-space search to analyze the structure of the flow table in existing networks, and finding the matching rules of the same type;
\item Integrating the analysis model with the dictionary tree using a tuple-space approach, and obtaining a rule matching strategy.
\end{itemize}

The rest of the paper is organized as follows. In Section II, we introduces challenges for flow tables of OpenFlow switches. We describes the problem of flow tables of SDN switches and presents a system model for flow tables in Section III. In Section IV, we realize the flow tables and design the structure of packets and flow tables. Section V presents the experimental results and evaluates the performance of the proposed matching model. Finally, conclusions are given in Section VI.

\vspace{0.3in}
\section{Challenges for the Flow Table of Openflow Switches}

In traditional network devices, the data forwarding of switches and routers depends on the MAC forwarding table on Layer 2 or the IP address routing table on Layer 3. The flow table of OpenFlow switches is the entry of network devices integrating other levels of OpenFlow networks. Each flow table of OpenFlow switches consists of three parts: header fields for matching packets (\textbf{Header Fields}), counters for counting the number of packets (\textbf{Counters}) and actions on how to process a matching packet (\textbf{Actions}). The OpenFlow header field is used to match the header of the packets received by switches. In OpenFlow v1.0, the header of the flow table contains twelve tuples. However, there are still following challenges for flow tables of OpenFlow switches.

\begin{enumerate}
  \item The number of matching fields of flow entries is increasing rapidly. OpenFlow v1.0 defines a 12-tuple matching field. OpenFlow v1.3 defines forty matching fields, including thirteen matching fields that the switches must support. It is an inevitable trend for networked big data to extend matching field types and increase the number of flow tables.
  \item The scale of flow tables grows exponentially. Along with the expansion of the network, the number of flow entries and the number of matching entries will continue to grow. OpenFlow v1.1 points out the scalability of flow tables by introducing a multilayer structure. However, the flexibility of multilayer flow tables is far beyond the capacity of data forwarding processing of traditional switching chip. It will add the complexity of designing hardware and software of OpenFlow switches.
  \item According to the application scenario of switches and the types of flow actions supported by switches, OpenFlow switches are classified into OpenFlow-only and OpenFlow-enabled, and the latter is also called OpenFlow-hybrid after OpenFlow v1.1. The former only supports OpenFlow protocol, while the latter considers the incompatibility between the OpenFlow switches and traditional switches when they are mixed. It can run the OpenFlow protocol and traditional Layer 2 or Layer 3 protocol stacks simultaneously. As a result, the latter is able to support \emph{NORMAL} action in the optional forward action in OpenFlow.
  \item After an OpenFlow switch receives network packets, the processing of 802.1d protocol in the flow will become an optional step in the process. When the OpenFlow switch receives a packet, it will match the entries with its local flow table according to the priority. As the result of matching, it takes the matching entry with the highest priority as the matching result based on the corresponding action to operate the packet. At the same time, once the matching is successful, the corresponding counter will update. If the packet cannot find a matching entry, it will forward to the controller.
\end{enumerate}

\section{The Structure of F-OpenFlow}

\subsection{Problem Description}
In recent years, the SDN has been widely used in the current experimental environments for network innovation, and it has broad application prospects as well as data centers. However, many critical technologies of the SDN have no mature solutions \cite{b15}. One of them is the update and design scheme of flow tables in switches.

The flow table of SDN switches contains a large number of matching rules. When a packet enters the flow table, it is matched in accordance with a specified search algorithm. Consequently, some matching algorithms, including linear and sequential search algorithms, will have a great effect on the pros and cons of SDN switches in the matching performance of data packets. The order of matching rules is limited by the search speed in the design level. Some linear and sequential rules in flow tables are analyzed, and a new pre-match method is formed by the hierarchical structure of data packets. Each packet, passing through flow tables in SDN switches, needs to be compared with the pre-model.

With the extension of OpenFlow applications, the scale of flow tables in OpenFlow switches is escalating, and the number of matching fields becomes larger and larger. The OpenFlow v1.1 \cite{b16} is proposed to compress the storage space and improve the lookup speed by a multilayer flow table and pipeline structure. The flow table of OpenFlow switches is divided into several word tables based on the protocol level of the header in the matching field. When the flow table is used for matching rules, the packet header is divided into multiple fields. We make a search operation in each field independently, use the method of sorting and grouping to find and integrate returned data, and return them to the flow table in OpenFlow switches for finding the rules of the \emph{Action}.

\subsection{The Structure of F-OpenFlow}

\begin{figure}[H]
\centering
\includegraphics[width=5in]{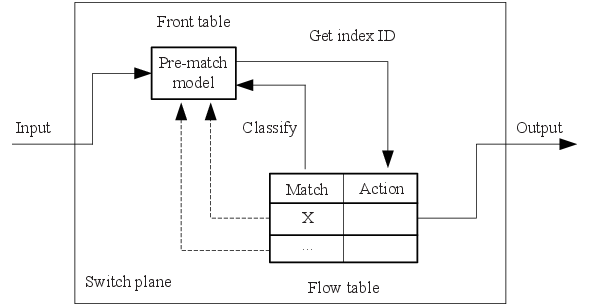}
\caption{The Structure of F-OpenFlow.}
\label{figure1}
\end{figure}

A flow entry is composed of six parts, including matching field, table item priority, counter, table item instruction, timeout period, and cookie. The matching field and the table item priority in the flow table of OpenFlow switches define a unique flow table entry. The matching field contains many matches, which cover most of the character sets in the link layer, network layer, and transport layer. The table item priority indicates the execution order of the flow entry. The table item instruction shows that the packet executes next instruction to match the table entry. The counter collects the statistics of the data flow, including the number and the length of matching the data flow. The devices in the OpenFlow network will no longer distinguish between routers and switches. In order to improve the efficiency of the flow table lookup, the current lookup uses the multilayer flow table and pipeline mode to obtain the corresponding operation. The OpenFlow pipeline of each switch contains multiple levels of flow tables that contain multiple flow entries. In this section, we present the structure of a matching model of the flow table in SDN switches, F-OpenFlow for short, as shown in Fig.\ref{figure1}.

Firstly, the data packets enter into the OpenFlow switch via a port of network devices. Secondly, by applying a specific analytical model, we gain a pre-match model for the corresponding packet inside the matching field. Thirdly, according to adopting the layered protocol, the packet data is able to be classified by the tuple-space lookup. Fourthly, we can obtain the corresponding front table model by adopting the results that are obtained in the third step. Finally, we get the index ID of the flow table to fetch an instruction from the action field in the table, and perform the instruction of action.

\section{Design Approach of F-OpenFlow}

\subsection{Implementation of Flow Tables}

The matching features of packets are as follows: Layer 1 is the ingress port (IngressPort) in switches. Layer 2 is made up of five parts which are the source MAC address (Ethersource), destination MAC address (Etherdst), Ethernet type (EtherType), VLAN tag (VLANid), and VLAN priority (VLANpriority). Layer 3 includes source IP (IPsrc), destination IP (IPdst), IP protocol field (IP proto), and IP service type (IP ToS bits). Layer 4 contains TCP/UDP source port number (TCP/UDP src port), and TCP/UDP destination port number (TCP/UDP dst port).

\begin{figure}[H]
\centering
\includegraphics[width=6.5in]{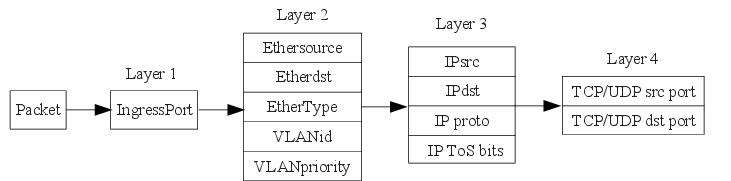}
\caption{Implementation process of the flow table in the F-OpenFlow.}
\label{figure2}
\end{figure}

Based on the concept of TCP/IP layers, we classify the matching fields of data packets. At the same time, we can gather the counts for the fields under classification phase. Fig.\ref{figure2} shows the implementation process of the flow table in the F-OpenFlow. The information of data packets is resolved, classified and counted. This categorization only considers the pre-match model of the field and ignores the specific value of the latter entry.

\subsection{Feature Structure of Data Packets}

Here are five data packets for analyzing the samples whose statistics are stored in a database table for comparison and classification, as shown in Tab.\ref{tab1}. It shows the relationship between the ID of packets and the number of matching fields in the corresponding layer.

\begin{table}[H]
\centering
\caption{Flow Characteristic Structure}
\begin{tabular}{|c|c|c|c|c|}
\hline
\textbf{Packet}& \textbf{Layer 1}& \textbf{Layer 2}& \textbf{Layer }3& \textbf{Layer 4}\\
\hline
Packet 1& 1& 2& 3& 1\\
Packet 2& 0& 3& 2& 1\\
Packet 3& 1& 1& 1& 1\\
Packet 4& 1& 3& 2& 1\\
Packet 5& 1& 2& 3& 1\\
\hline
\end{tabular}
\label{tab1}
\end{table}

From Tab.\ref{tab1}, we can see that Packet 1 contains one first-level matching field, two second-level matching fields, third third-level matching fields, and one fourth-level matching field. The structure of Packet 5 is similar to that of Packet 1. Obviously, we can classify the data according to the number of hierarchical statistics. For example, the structural model of Packet 1 and Packet 5 can be classified the two as a homogeneous structure. When the pre-matching model matches the corresponding structure, the number of matches in the corresponding type of flow tables can be found quickly.

\subsection{Feature Structure of Flow Table Rule}

\begin{figure}[H]
\centering
\includegraphics[width=5in]{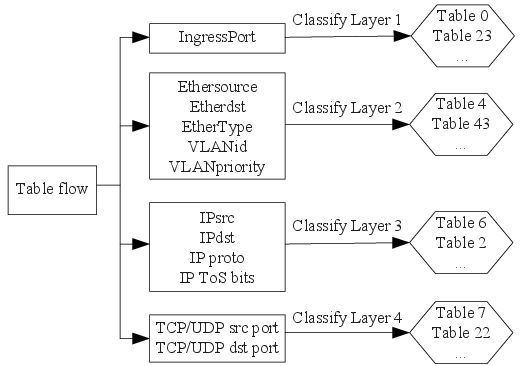}
\caption{F-OpenFlow rules for classifying flow tables.}
\label{figure3}
\end{figure}

Fig.\ref{figure3} shows rules for classifying flow tables. For each table, we count the matching fields based on the previous entries, and ignore other parameters. Assume that there are sixty tables,  using the above classification method, we can find that there are seven tables in the first layer, twenty-one tables in the second layer, nineteen tables in the third layer, and thirteen tables in the fourth layer for meeting the requirement.

At the same time, we gain the index ID of each table in order to find the root, as shown in Tab.\ref{tab2}. In this way, it is helpful to locate the approximate position of all matching fields in the flow table quickly and precisely.

\begin{table}[H]
\centering
\caption{Statistics on all tables}
\begin{tabular}{|c|c|c|c|c|}
\hline
\textbf{Table}& \textbf{Layer 1}& \textbf{Layer 2}& \textbf{Layer 3}& \textbf{Layer 4}\\
\hline
Table number& 7& 21& 19& 13\\
Table ID& 0, 1, 3, 5, ...& 4, 9, 11, 13, ...& 6, 12, 14, ...& 7, 22, 24, ...\\
\hline
\end{tabular}
\label{tab2}
\end{table}

\subsection{Classification Algorithm}

We categorize data packets on the concept of TCP/IP layer and obtain layered packets, and it shortens the matching range of packets in switches. The specific methods to achieve the following: (1) getting all matching fields in the data packet; (2) omitting specific parameters after the field and only matching the first item; (3) classifying statistics and getting fuzzy results; (4) converting a corresponding numeric string into resulting statistics. We implement this method with the code shown in Algorithm \ref{alg1}.

\begin{algorithm}[H]
\caption{\textbf{:}~~Packet classification.}
\label{alg1}
\begin{algorithmic}[1]
\REQUIRE~Data packets.
\ENSURE~Layered packets.
\FOR {$Packge$ in $TCP/IP$}
\IF {$Packget.String=TCP/IP.String$}
\STATE $Packget.layer=TCP/IP.layer$
\STATE $Packget.Number= Packget.Number+1$
\ENDIF
\ENDFOR
\end{algorithmic}
\end{algorithm}

The core idea of OpenFlow flow tables passing through F-OpenFlow transformation is: (1) obtaining the results of both sides of the classification algorithm;
(2) comparing and confirming a rough rule model of flow tables; (3) finding the specific model ID in the front table, and getting its corresponding rules ID
in original tables; (4) according to the ID of the original flow table, matching all the operation of Action. The code is shown in Algorithm \ref{alg2}.

\begin{algorithm}[H]
\caption{\textbf{:}~~Flow table term hierarchy.}
\label{alg2}
\begin{algorithmic}[1]
\REQUIRE~OpenFlow flow table.
\ENSURE~F-OpenFlow fuzzy hierarchical model.
\FOR {$Flow$ in $TCP/IP$}
\IF {$Flow.Table= TCP/IP.String$}
\STATE $Flow.layer=TCP/IP.layer$
\STATE $Flow.Number= Packget.Number+1$
\STATE $Flow.Table_id.add(Flow.id)$
\ENDIF
\ENDFOR
\end{algorithmic}
\end{algorithm}

The matching becomes the principle of fuzzy matching after getting the packet feature code and the fuzzy classification structure of the flow table. There are two reference elements during the period: the hit rate and the tuple length. Where A represents the data package and B represents the flow.

The fuzzy matching code between the data packet and the two features of the flow table entries is given in Algorithm \ref{alg3}.

\begin{algorithm}[H]
\caption{\textbf{:}~~F-OpenFlow.}
\label{alg3}
\begin{algorithmic}[1]
\REQUIRE~Data packets and stream tables correspond to the F-OpenFlow model.
\ENSURE~Hit rate and length of tuple.
\FOR {$A,B$ in $F$-$OpenFlow$}
\IF {$A.layer= Flow.layer.number$}
\FOR {$A.layer.Match in Flow.layer.Match$}
\IF {$A.layer.Mathch==Flow.layer.Match$}
\STATE $A.Hit.number++$
\ENDIF
\ENDFOR
\ENDIF
\STATE $A.Hit=A.Hit.number/A.layer.Length$
\ENDFOR
\end{algorithmic}
\end{algorithm}

\section{Experimental Result}
In order to ensure the efficiency and practicability of the proposed architecture, we have built a data model which is similar with the flow table in the database. By using Java language, we design the flow table architecture for optimizing OpenFlow.

By using OpenFlow 1.0, we first find and match twelve elements, and then we carry on the data consolidation according to the data warehouse theory. The basic database is constructed to be used in real-time data, and it is the basis for establishing analysis platform.

\subsection{Hit Rate}
We use the proposed F-OpenFlow and the original flow table structure to carry on the related data in our test experiments. We compare the two groups of data packets through two different designs.

\begin{figure}[H]
\centering
\includegraphics[width=3.22in]{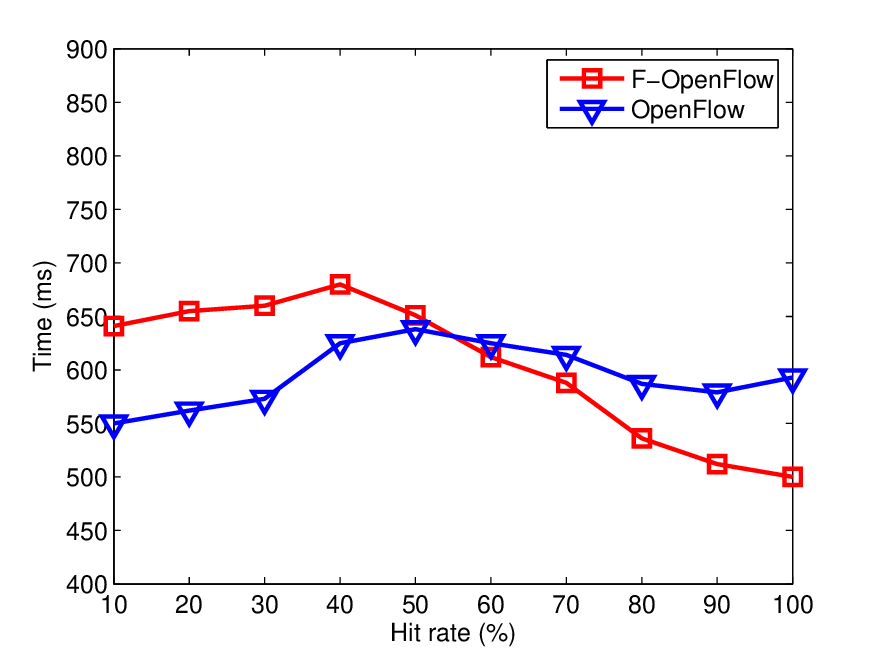}
\includegraphics[width=3.22in]{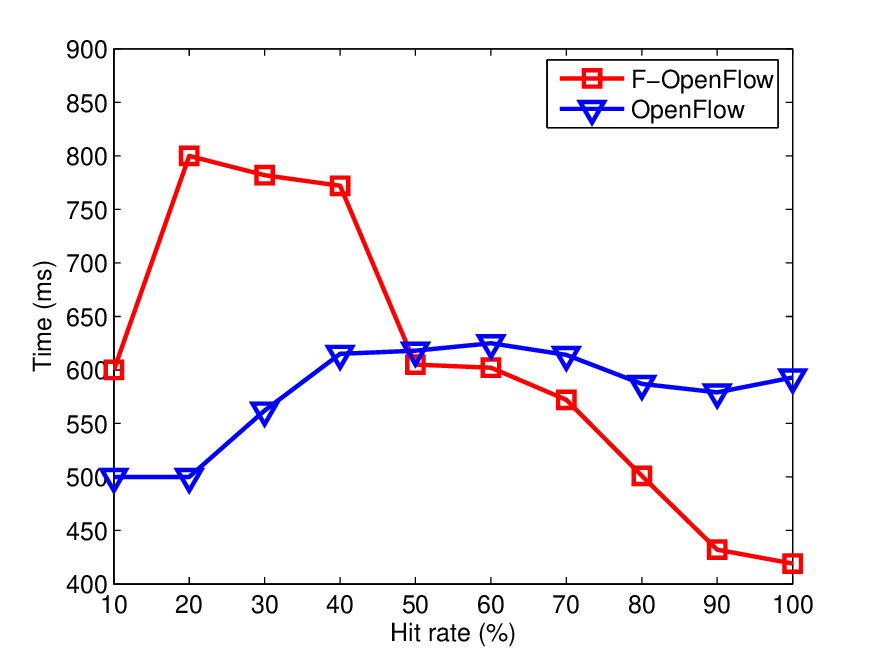}
\caption{Hit rate vs. time (left: 10 tuples; right: 12 tuples).}
\label{figure4}
\end{figure}

Fig.\ref{figure4} shows the impact of the time (in ms) to enter the flow table from a single data flow to the Action instruction on the hit rate of the data flow to the field in the forward model of flow tables.

When the hit rate is 10$\%$, there is a higher time. Compared with the original structure, the F-OpenFlow has more steps, so there is a higher time. When hit rate is 20$\%$, 30$\%$, and 40$\%$, respectively, the time has declined while the effect is not obvious. But the gap between the two structures is constantly narrowing. When the hit rate is equal to 50$\%$, and there is a critical point that indicates that the original structure and the proposed F-OpenFlow are mutually balanced. When the hit rate is more than 50$\%$, the superiority is quite obvious.

From Fig.\ref{figure4}, it can be seen that when the contrast value of the matching rate between the data flow and the field in the pre-model flow table is very low, the flow table structure of the proposed F-OpenFlow does not have the original structure quickly. However, with the improvement of the hit rate, it can be seen that there are obvious changes in the structure design of the proposed F-OpenFlow. When the hit rate is close to 50$\%$, the time required for both structures is very close. When the hit rate is equal to 70$\%$, the F-OpenFlow is significantly better than the original design.

\subsection{Tuple length}

\begin{figure}[H]
\centering
\includegraphics[width=3.22in]{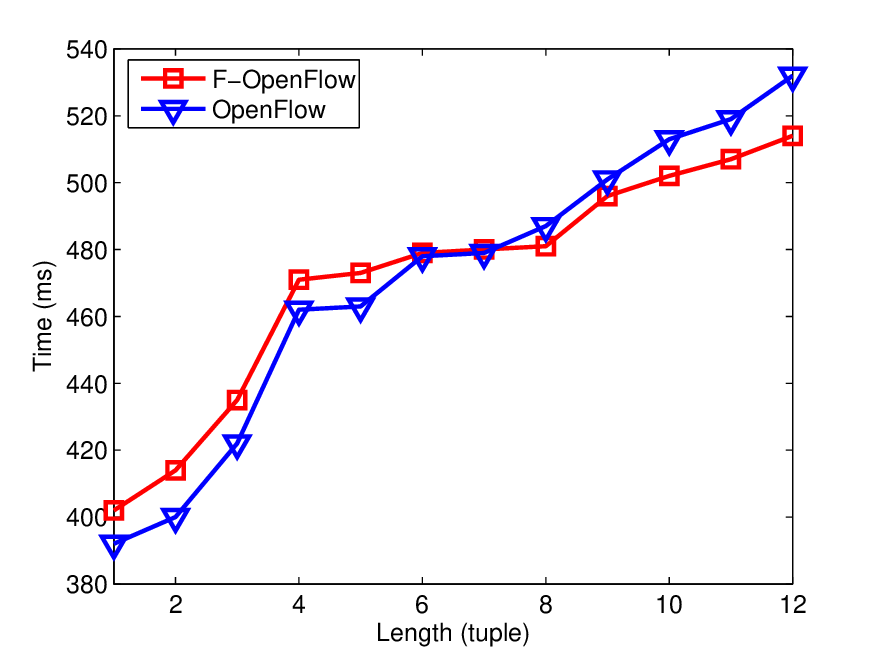}
\includegraphics[width=3.22in]{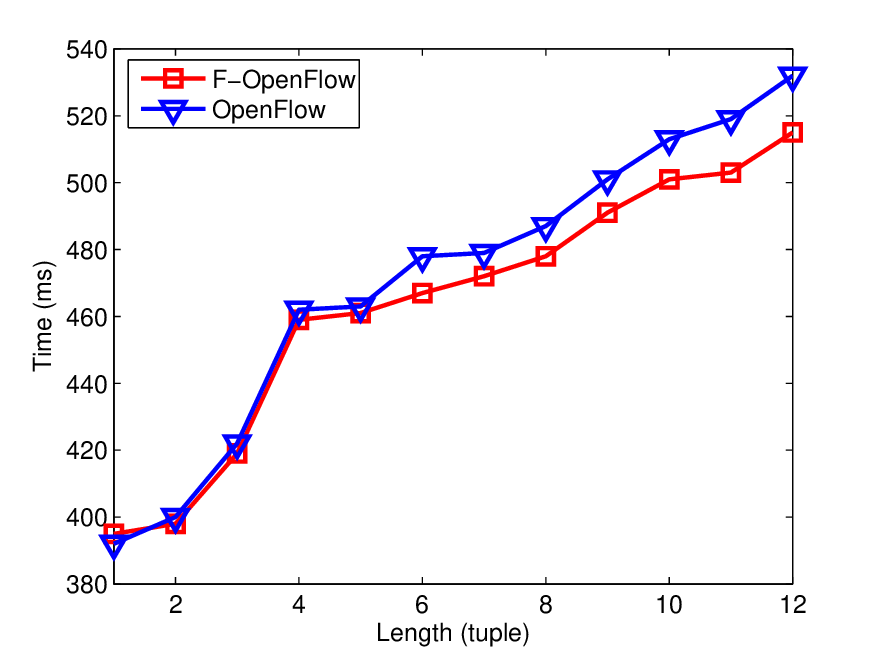}
\caption{Tuple length vs. time (left: average hit rate; right: hit rate is 100$\%$).}
\label{figure5}
\end{figure}

Fig.\ref{figure5} shows the impact of the time to enter the flow table from a single data flow to the Action instruction on the tuple length of the matching field. The single data flow gets into the flow table for getting the time to get the Action command (in ms), and the matching field is the one that is in the flow table of the proposed OpenFlow.

From Fig.\ref{figure5}, it can be seen that for a tuple length, the original structure is shorter than the structure of the F-OpenFlow, because of the very short length. This phenomenon is not static. The time for the two structures increases on 2, 3, 4, 5, and 6 tuple lengths. The growth rate of the optimized structure of the F-OpenFlow is lower than that of the original structure. When the length of the tuple exceeds seven, the time of the optimizing structure of the F-OpenFlow is less than that of the original structure.

We have simulated twelve test cases of tuple length. We found that when the length is relatively short and the inner tuple capacity is small, the proposed F-OpenFlow does not match the OpenFlow fast. But with the increase of the tuple length, we can see that the F-OpenFlow matching structure design has a relatively good development trend in the experimental results. When the tuple length increases to eight, the advantages and disadvantages of F-OpenFlow and OpenFlow have obvious difference. Accordingly, it can directly verify that the structural design of the F-OpenFlow has the advantage of shortening the time of the long data flow.

\subsection{Frequency}

\begin{figure}[H]
\centering
\includegraphics[width=3.22in]{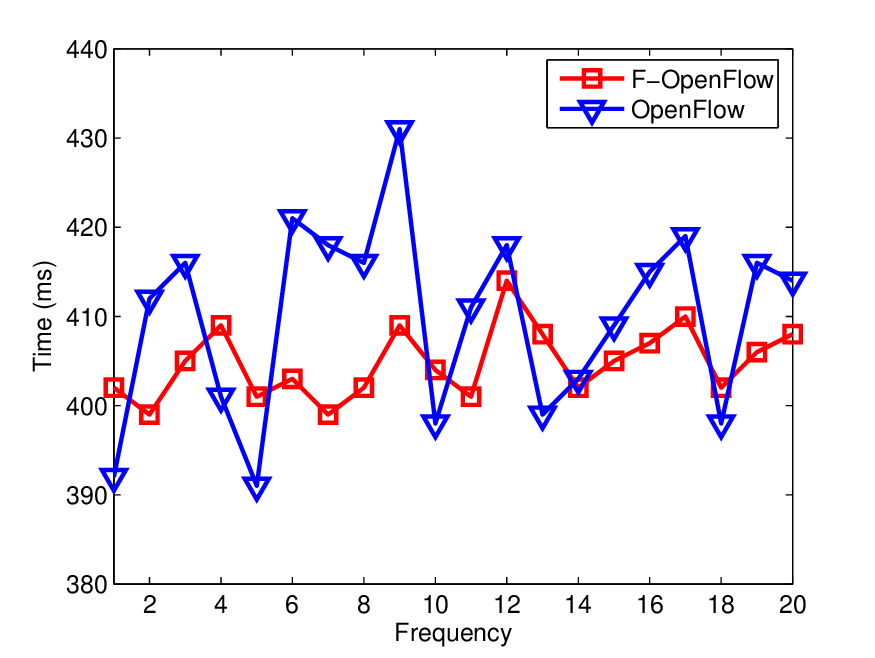}
\includegraphics[width=3.22in]{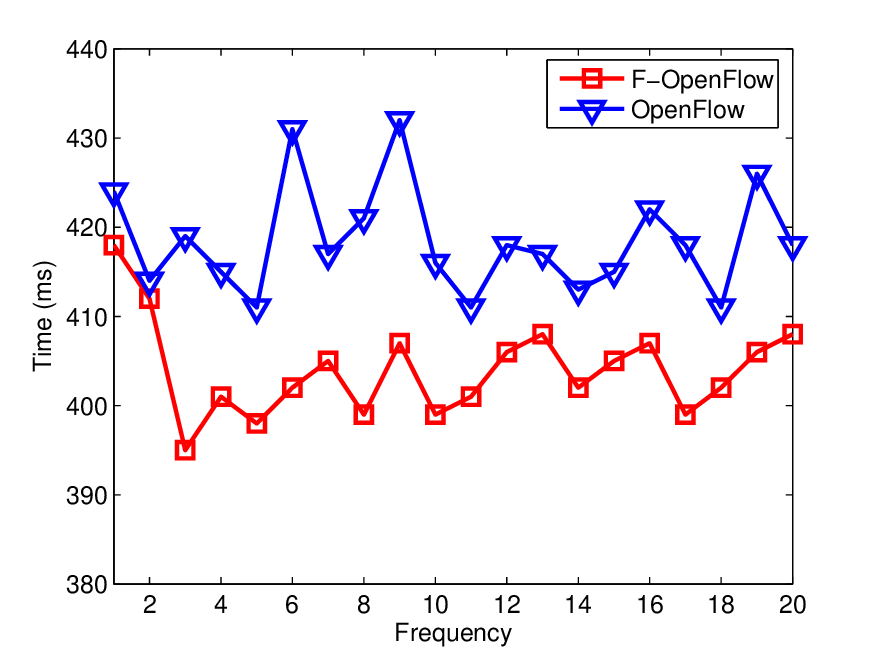}
\caption{Frequency vs. time (left: hit rate is 50$\%$ and 8 tuples; right: hit rate is 100$\%$ and 12 tuples).}
\label{figure6}
\end{figure}

Fig.\ref{figure6} shows the impact of the time (in ms) that a single data flow enters the flow table until it gets the Action instruction on the number of repeated tests for the same data flow table. From the above two tests we have chosen the middle segment of a group of data packets, and the data packets with two dimensions of hit rate and tuple length are the 50$\%$ of length and the eight elements. The difference between the two can be seen in terms of the data. The stability of the new proposed structure of the F-OpenFlow is better than that of the original design.

\section{Conclusions}
For the wisdom of the networked big data in this direction combined with the scalability problem of SDNs, the performance of SDN switch directly affects the speed of information exchange and the management process. The proposed F-OpenFlow structure is different from the OpenFlow. It is optimized on the basis of the original one. The purpose is to solve the problem of packet matching speed. The core idea of the F-OpenFlow is to give priority to the analysis of the content structure of flow table rules, and to establish a simple model of the front table. In the proposed F-OpenFlow structure, the packet matching can quickly locate the specific Action instruction of the flow table. In case of the packet length is longer, F-OpenFlow matching is better than OpenFlow matching. The future work will also take advantage of the previous historical data of matching. It will be more accurate to analyze the relationship between the location of the message and the flow table. In this way, we can improve the hit rate and matching speed of the flow table.

\vspace{0.3in}
\section{Acknowledgements}
This work was supported in part by Natural Science Foundation of China (Grant No. 61572191 and 61402170).


\end{document}